\documentclass[12pt]{article}
\usepackage{amsmath,amssymb,bm,epsfig}
\usepackage{amsfonts,amsbsy,cite}
\usepackage{graphicx}
\usepackage{ulem}
\usepackage{color}
\usepackage{subfigure}
\usepackage{mathrsfs}
\usepackage{amsfonts}
\usepackage{latexsym}
\usepackage{epstopdf}
\usepackage{psfrag}

\renewcommand{\thefootnote}{\fnsymbol{footnote}}
\input epsf

\def\sla#1{\rlap{\kern .15em /}#1}

\newcommand{\calP}{{\cal P}}
\newcommand{\calR}{{\cal R}}

\newcommand{\calI}{{\cal I}}
\newcommand{\calC}{{\cal C}}

\begin{document}

\title{
\begin{flushright}
\begin{minipage}{0.2\linewidth}
\normalsize
CTPU-15-14 \\
EPHOU-15-013 \\
WUF-HEP-15-017 \\*[50pt]
\end{minipage}
\end{flushright}
{\Large \bf 
Axion inflation with cross-correlated axion isocurvature perturbations
\\*[20pt]}}

\author{Kenji~Kadota$^{1}$\footnote{
E-mail address: kadota@ibs.re.kr},\ \ 
Tatsuo~Kobayashi$^{2}$\footnote{
E-mail address:  kobayashi@particle.sci.hokudai.ac.jp}, \ and \ 
Hajime~Otsuka$^{3}$\footnote{
E-mail address: hajime.13.gologo@akane.waseda.jp
}\\*[20pt]
$^1${\it \normalsize 
Center for Theoretical Physics of the Universe, Institute for Basic Science, 
Daejeon 305-811, Korea} \\
$^2${\it \normalsize 
Department of Physics, Hokkaido University, Sapporo 060-0810, Japan} \\
$^3${\it \normalsize 
Department of Physics, Waseda University, 
Tokyo 169-8555, Japan}\\*[50pt]}

\date{
\centerline{\small \bf Abstract}
\begin{minipage}{0.9\linewidth}
\medskip 
\medskip
\small
We study the inflation scenarios, in the framework of superstring theory, where the inflaton is an axion producing the adiabatic curvature perturbations while there exists another light axion producing the isocurvature perturbations.
We discuss how the non-trivial couplings among string axions can generically arise, and calculate the consequent cross-correlations between the adiabatic and isocurvature modes through concrete examples. Based on the Planck analysis on the generally correlated isocurvature perturbations, we show that there is a preference for the existence of the correlated isocurvature modes for the axion monodromy inflation while the natural inflation disfavors such isocurvature modes.
\end{minipage}
}

\begin{titlepage}
\maketitle
\thispagestyle{empty}
\clearpage
\tableofcontents
\thispagestyle{empty}
\end{titlepage}

\renewcommand{\thefootnote}{\arabic{footnote}}
\setcounter{footnote}{0}
\vspace{35pt}

\section{Introduction} 
The recent Cosmic Microwave Background (CMB) data is still consistent with the simple $\Lambda$CDM model with a nearly scale-invariant, adiabatic and Gaussian power spectrum which can well be represented by the single-field 
slow-roll inflation models~\cite{ben,valv,Ade:2015lrj}.
The forthcoming cosmological data with even better precision however could reveal the potential deviations from such pure adiabatic perturbations, and it would be worth exploring the possibilities for the non-adiabatic perturbations in existence of correlations among the adiabatic and non-adiabatic modes along with their indications for the early Universe dynamics in the fundamental physics models. 
In this paper, we study the mixture of the adiabatic mode and cold dark matter (axion) isocurvature mode taking account of their possible cross-correlations~\cite{amen,pol,lang,cro,kur,mack,kadojinn,gon,bucher} through the concrete models based on superstring theory. 
The recent Planck analysis studied the generally correlated isocurvature perturbations. The robust parameter estimation, without significantly affecting the bounds on the conventional ($\Lambda$CDM) cosmological parameters even with the inclusion of isocurvature modes, was not previously realized due to strong degeneracies among the parameters involving the isocurvature perturbations which the previous CMB data sets suffered from and the Planck data could greatly reduce  \cite{Ade:2015lrj}.
With such a precise cosmological parameter estimation including the correlated isocurvature perturbations at hand, it would be intriguing to explore the indications of the generally correlated isocurvature perturbations for the early Universe phenomena, and we in this paper aim to study the presumably ubiquitous light degrees of freedom in the early Universe through their isocurvature fluctuations. 

In superstring theory, the higher-dimensional form fields 
predict a number of light degrees of freedom represented by the axions in addition to the QCD axion~\cite{Peccei:1977hh}. 
These axions are associated with the internal cycles of the extra-dimensional manifold. 
While, at the perturbative level, the axion potential is protected by the gauge symmetry in string theory, the non-perturbative effects can break the continuous gauge symmetry leading to the discrete one and generate the axion potential. It is thus expected that the axion potential is well controlled by the residual discrete symmetry and the mass scale of axions depends on the non-perturbative 
effects \cite{Blumenhagen:2006ci,Ibanez:2012zz,Baumann}. 
We, in this paper, focus on the single-field axion inflation models in coexistence of an isocurvature perturbation due to another light axion. 

Although the axion inflation is often considered as the single-field inflation, the axion potential, in general, has the axionic mixing due to the moduli-mixing 
gauge kinetic function. As concrete examples, we discuss the natural inflation~\cite{Freese:1990rb} in Sec.~\ref{sec:2} and axion monodromy 
inflation~\cite{McAllister:2008hb,Silverstein:2008sg} in Sec.~\ref{sec:3} where the cross-correlated isocurvature perturbations can arise due to such axionic mixings. 
In these illustrative examples, the adiabatic curvature perturbations are dominantly sourced by the axion-inflaton whereas the isocurvature perturbations originate from the fluctuations of the light axion (different from the heavy axion inducing the inflation). 
The mixing of the string axions arises from the non-perturbative effects in the sinusoidal form, and the consequent cross-correlations between the adiabatic and isocurvature modes are studied.
We conclude our discussions in Sec.~\ref{sec:con}.



\section{Natural inflation with sinusoidal correction}
\label{sec:2}
The natural inflation is among the simplest axion inflation models ~\cite{Freese:1990rb} which can be 
constructed in the field theory as well as superstring theory. 
Although, at the perturbative level, the axion potential is not generated due to the gauge symmetry in string theory, the non-perturbative effects in a hidden gauge sector 
can generate the axion potential terms. 
Especially, when the gauginos ($\lambda$) of the hidden gauge group condensate at a certain energy scale~\cite{Ferrara:1982qs}, 
the superpotential can be generated in the four-dimensional ($4$D) ${\cal N}=1$ supersymmetry, 
\begin{align}
W\simeq \langle \lambda \lambda \rangle \simeq Ae^{-aT},
\end{align} 
where $A={\cal O}(1)$ and $a=24\pi^2/b_0$ with $b_0$ being the one-loop beta-function coefficient.\footnote{Here and in what follows, we 
employ the reduced Planck units, $M_{\rm Pl}=2.4\times 10^{18}\,{\rm GeV}=1$.} 
We consider the scenarios where the size of gauge coupling is determined 
by the real part of modulus field, $T$, which is typical for heterotic string theory, type I string theory 
and type II string theory with D-branes along the single cycle (see for reviews, e.g., Refs.~\cite{Blumenhagen:2006ci,Ibanez:2012zz}).
By fixing the real part of moduli, for instance through another non-perturbative 
effect, we can obtain the effective inflaton potential for the imaginary part of modulus (axion),
\begin{align}
V_{\rm inf}=\Lambda_1^4 \left( 1-{\rm cos}\frac{\phi}{f} \right),
\end{align} 
with $\phi$ and $f$ being the axion-inflaton and its decay constant. 
The conventional natural inflation model, in view of the recent Planck data, requires the trans-Planckian axion decay constant, $f>5$, even though its construction requires some care because the fundamental axion decay constant obtained after the dimensional reduction is typically much smaller than the 
Planck scale \cite{Choi:1985je}. 
The gauge couplings in the visible and hidden sectors, in general, depend on the linear combination 
of moduli fields through the gauge threshold correction and non-trivial brane configuration. 
For example, in type II string theory, the D-branes wrap the internal cycle of six extra-dimensional 
manifold and then the volume of this internal cycle is determined by the linear combination of 
moduli fields $T_i$ where the number of moduli $T_i$ is determined by the 
topology of the extra-dimensional manifold ~\cite{Blumenhagen:2006ci,Ibanez:2012zz}.
Thus, the gauge coupling on Dp-branes is represented by 
the linear combination of them,
\begin{align}
\langle c_iT^i\rangle =\frac{1}{g^2}, 
\end{align} 
where $c_{i}$ are constant. 
Furthermore, if we consider the one-loop corrections for the gauge coupling, 
the superpotential also depends on the linear combination of moduli fields $T$ and $T^\prime$,
\begin{align}
W=Ae^{-aT-dT^\prime},
\label{eq:threshold}
\end{align} 
where $A={\cal O}(1)$, $a=24\pi^2/b_0$ and $d=24\pi^2/b_0 \times b/48\pi$ with $b$ being 
the one-loop beta-function coefficient determined by massive modes \cite{Lust:2003ky}. The axion decay constant for the modulus $T^\prime$ here can be enhanced by 
the one-loop effect~\cite{Abe:2014pwa,Abe:2014xja}. Indeed, there are several scenarios to enhance 
the axion decay constant based on the moduli-mixing in the gauge kinetic function such as the alignment mechanism~\cite{Kim:2004rp}, N-flation~\cite{Dimopoulos:2005ac}, 
kinetic mixing~\cite{Bachlechner:2014hsa}, the threshold correction~\cite{Abe:2014pwa,Abe:2014xja} and 
the flux-induced enhancements~\cite{Hebecker:2015rya}.

Since there exist, in general, ubiquitous axion fields in string theory, one can also expect that there are moduli-dependent correction terms in the potential,
\begin{align}
V_{\rm int}=\Lambda_2^4 \left( 1-{\rm cos}\left(\frac{\phi}{g_1} +\frac{\chi}{g_2}\right) \right),
\label{eq:infponat2}
\end{align} 
where $\phi$ and $ \chi$ represent an axion-inflaton and another light axion field. 
For the notational brevity, we in the following define the parameters 
\begin{align}
\sigma =\frac{\phi}{f},\,\,\,
\psi =\frac{\phi}{g_1},\,\,\,
\theta =\frac{\chi}{g_2},
\end{align} 
so that the total potential can be written as
\begin{align}
V=\Lambda_1^4 (1-\cos\sigma)
+
\Lambda_2^4
\left( 1-
\cos\left(\psi+\theta\right)
\right).
\label{eq:infponat3}
\end{align} 
We hereafter focus on the scenarios where the adiabatic perturbations are dominantly sourced by the axion-inflaton fluctuation $\delta \phi$ 
and the additional axion fluctuation $\delta \chi$ leads to the isocurvature perturbations. 

\begin{figure}
\begin{center}
  \epsfxsize = 0.48\textwidth
    \psfrag{nRR}[B][B][1][0]{$n_s$}
  \psfrag{rrr}[B][B][1][0]{$r$}
  \includegraphics[scale=0.4]{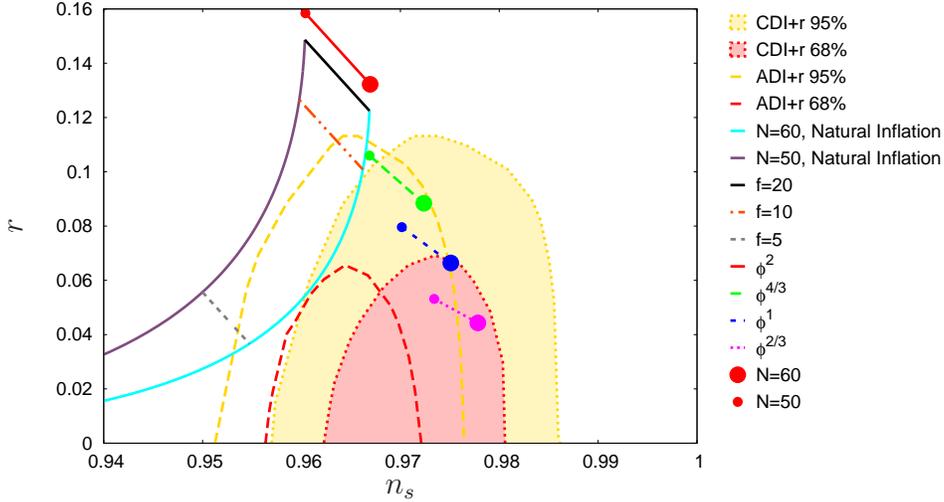}
\end{center}
\caption{$68\%$ and $95\%$ confidence level constraints on the adiabatic spectral index and tensor to scalar ratio from Planck \cite{Ade:2015lrj}.
The filled contours are for generally-correlated adiabatic and CDM (axion) isocurvature modes. 
The unfilled dashed contours are for the pure adiabatic model without the isocurvature 
perturbations.}
\label{nsvsr}
\end{figure}

Before discussing the cosmological perturbations and their indication for the model discrimination, we show an allowed parameter region for the spectral tilt of the adiabatic perturbations and the tensor-to-scalar ratio in both pure adiabatic (ADI) model and generally-correlated ADI + cold dark matter isocurvature (CDI) model.
Fig.~\ref{nsvsr} shows that, for the natural inflation with isocurvature perturbations, the inclusion of a cross-correlated isocurvature mode tightens the constraints on the axion decay constant, $5<f<10$, and the e-folding number, $60<N$. 
On the other hand, the axion monodromy inflation, to be discussed in the next section, 
except for the quadratic one can be better fitted by the Planck data by including the cross-correlated isocurvature mode. 

The degeneracies among the parameters involving the correlated isocurvature perturbations result in the shift in the best-fit parameters compared with those in the pure adiabatic model, even though the strong degeneracies such as that between the isocurvature perturbation amplitude and adiabatic perturbation spectral index which WMAP data had greatly suffered from reduced significantly in Planck $TT$ + polarization data \cite{ben,valv}. The constraints on $r$ however turn out not to be significantly affected by the inclusion of the cross-correlated isocurvature modes partly because the Planck data including the polarization already gives sufficiently tight constraints on the isocurvature and tensor modes \cite{Ade:2015lrj}. In the rest of the paper, we for simplicity do not consider the significant tensor contribution and we adopt the Planck likelihood analysis results without including $r$ in the following discussions.

We now discuss the cosmic perturbations for the axion fields, starting with the brief discussions for the conventional curvature and isocurvature perturbations to set up our notations followed by the exploration on their cross-correlations along with their indication for the string inflation model building \cite{Kadota:2014hpa}. 
The curvature and isocurvature perturbations in our scenarios are \cite{ewanlyth, Kadota:2014hpa}
\begin{align}
{\cal R} & = -\frac{H}{\dot\phi_0} \delta \phi \, ,
\\
{\cal I} & = 2\frac{\Omega_a}{\Omega_m} \frac{\delta \theta}{\theta_0} \, ,
\end{align} 
with $\Omega_a$ and $\Omega_m$ being the axion and matter densities with respect to the critical density. The factor $\Omega_a/\Omega_m$ appears here because we are interested in the isocurvature perturbations between the radiation and the non-relativistic matter, and the non-adiabatic fluctuations arise solely from an axion which contributes to the total matter density with the fraction $\Omega_a/\Omega_m$.
The dot denotes the time derivative and the subscript $_0$ represents the background field values during the inflation and, in the following, 
we omit $\delta$ representing the fluctuations for the notational brevity when it is clear from the context. 

The corresponding power spectra are given by~\cite{ewanlyth}
\begin{align}
{\cal P}_{\cal R} &=\left( \frac{H}{\dot\phi_0} \right)^2 {\cal P}_\phi = 
\nonumber\\
& = \left( \frac{H}{2\pi} \right)^2 \left( \frac{H}{\dot\phi_0} \right)^2 \left( \frac{k}{aH} \right)^{3-2\nu_\phi} 2^{2\nu_\phi-3} \left[ \frac{\Gamma(\nu_\phi)}{\Gamma(3/2)} \right]^2, 
\nonumber\\
{\cal P}_{\cal I}  &=
\left(\frac{\Omega_a}{\Omega_m}  \right)^2
\left( \frac{2}{\theta_0} \right)^2 
 {\cal P}_\theta =
\nonumber\\
& = \left(\frac{\Omega_a}{\Omega_m}  \right)^2
\left( \frac{H}{2\pi} \right)^2 
\left( \frac{2}{g_1\theta_0} \right)^2 \left( \frac{k}{aH} \right)^{3-2\nu_\theta}  2^{2\nu_\theta-3} \left[ \frac{\Gamma(\nu_\theta)}{\Gamma(3/2)} \right]^2, 
\label{eq:PRPI}
\end{align}
in terms of $\nu_{\phi(\theta)}=\sqrt{9/4-m^2_{\phi(\theta)}/H^2}$ with
\begin{align}
m_{\phi}^2&= \frac{\Lambda_1^4}{f^2}\cos \sigma_0,
\nonumber\\
m_{\theta}^2&= \frac{\Lambda_2^4}{g_2^2} \cos (\psi_0+\theta_0),
\nonumber\\
H^2&=\frac{V}{3}\simeq \frac{\Lambda_1^4}{6} \left( \frac{\phi_0}{f} \right)^2.
\end{align} 

\begin{figure}[htb!]
\begin{center}    
      \psfrag{nRR}[B][B][1][0]{$n_s$}
  \psfrag{PRI/Sqrt[PRR*PII]}[B][B][1][0]{$\calP_{\calC}/\sqrt{\calP_\calR\calP_\calI}$}
    \includegraphics[scale=0.4]{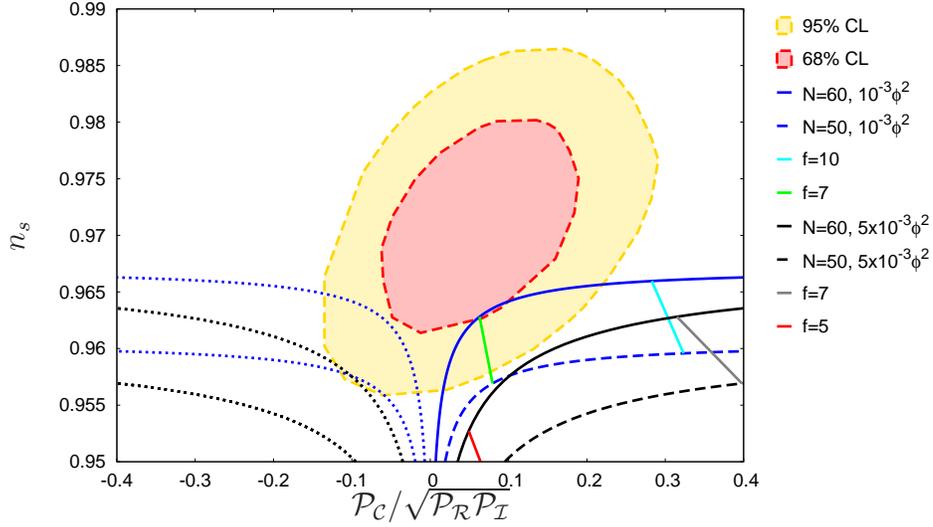} 
  \end{center}  
\caption{$\calP_{\calC}/\sqrt{\calP_{R}\calP_{I}}$ and the adiabatic spectral index $n_{s}$ for the natural inflation with sinusoidal correction (68 \% and 95 \% CL contours are from Planck \cite{Ade:2015lrj}). $\calP_{C}/\sqrt{\calP_{\calR}\calP_{\calI}} =c\times \phi_0^2$ for $c=10^{-3}, 5\times 10^{-3}$ are shown for varying $N$ and 
$f$ (the labels are in units of the reduced Planck mass). 
The anti-correlation cases (for $c=-10^{-3}, -5\times 10^{-3}$) are also shown with the dotted curves.}

\label{crosscnat}
\end{figure}

The cross-correlation between the curvature and isocurvature perturbations can be obtained using the in-in formalism~\cite{Weinberg:2005vy}, 
and the relevant interaction term in the Hamiltonian at the quadratic order, $\Lambda_2^4 \cos (\psi_0+\theta_0) \delta \psi  \delta \theta$, 
leads to the isocurvature cross-correlation power spectrum~\cite{Weinberg:2005vy,Kadota:2014hpa},
\begin{align}
{\cal P}_{\cal C}  &= - \frac{\pi}{2}\frac{\Lambda_2^4}{g_1 g_2 H^2} \cos (\psi_0 +\theta_0)\Re \left[ i\int_0^\infty \frac{dx}{x} H_{\nu_\phi}^{(2)}(x) H_{\nu_\theta}^{(2)}(x) \right] 
\sqrt{{\cal P}_{\cal R}{\cal P}_{\cal I}}
\nonumber\\
&\sim 
4.2 \left( \frac{1}{g_1 g_2} \right)
\left(
\frac{\Lambda_2}{\Lambda_1}
\right)^4
\left(
\frac{f}{\phi_0}
\right)^2
\cos(\psi_0+\theta_0)
\sqrt{{\cal P}_{\cal R}{\cal P}_{\cal I}},
\label{eq:PC}
\end{align} 
where the numerical integral of the Hankel function gives a factor 
$\sim -0.45$ and an order of the Hankel function is taken as 
$\nu_{\phi(\theta)}=\sqrt{9/4-m^2_{\phi(\theta)}/H^2}$.\footnote{This integral can be evacuated, at the leading order, at an arbitrary value of 
$x$ as long as $e^{-1/\xi} < x < 1$ with $\xi$ being the typical size of the slow-roll parameter \cite{gongsp,ks03}.} 
We plot the following cross correlation parameter
\begin{align}
\beta_C \equiv 
\frac{{\cal P}_{\cal C}}{\sqrt{{\cal P}_{\cal R}{\cal P}_{\cal I}}}
  \sim 4.2 \left( \frac{1}{g_1 g_2} \right)
\left(
\frac{\Lambda_2}{\Lambda_1}
\right)^4
\left(
\frac{f}{\phi_0}
\right)^2 \cos(\psi_0+\theta_0)
\simeq 
4.2 \left( \frac{1}{g_1g_2} \right)
\left(
\frac{\Lambda_2^4}{A_S}
\right) \cos(\psi_0+\theta_0)
\frac{\phi_0^2}{96\pi^2},
\end{align} 
by varying the axion decay constant $f$ from $1$ to $20$  and the e-folding number $N$ 
from $50$ to $60$ in Fig.~\ref{crosscnat}. 
Here, the prefactor of $\phi_0^2$ is set to be of order $0.001$ and 
the power spectrum of the adiabatic curvature perturbations  
\begin{align}
  {\cal P}_{\cal R}=A_Sk^{n_{s}-1},
  \label{askn}
\end{align} 
is fixed to be $A_S\simeq H^2/(8\pi^2 \epsilon) \simeq 2.2 \times 10^{-9}$ with $\epsilon\simeq 2/\phi_0^2$ being the slow-roll 
parameter at the pivot scale $k_{\ast}=0.05\,{\rm Mpc}^{-1}$. 
This figure also shows the Planck likelihood contours including the polarization data which greatly improve the constraints on the isocurvature perturbations compared with the WMAP results \cite{ben,valv}. The high-$l$ ($l\geq 30$) $TE, EE$ data turn out to drive the isocurvature cross-correlation towards a smaller value and disfavor the negative cross-correlations which would be allowed otherwise with the high-$l$ $TT$ data \cite{Ade:2015lrj}.  
We can find that the coefficient $c$ in $\beta_{\calC}=c\,\phi_0^2$ has to
 be of order less than $10^{-3}$ to be within $2$ sigma 
and the axion decay constant $f$ is constrained to the range between $5$ and $10$. 
The cross correlation parameter $\beta_{\calC}$ is constrained to be $-0.1\lesssim \beta_{\calC} \lesssim 0.3$, or, in terms of the parameters in the sinusoidal correction term ~(Eq.~(\ref{eq:infponat2})), 
to be within 
\begin{align}
-0.1 \lesssim 4.2 \left( \frac{1}{g_1g_2} \right)
\left(
\frac{\Lambda_2^4}{2.2\times 10^{-9}}
\right) \cos(\psi_0+\theta_0)
\frac{\phi_0^2}{96\pi^2} 
\lesssim 0.3.
\label{eq:betac1nat}
\end{align}

Moreover, the following conditions are taken into account to justify our calculations:
\begin{description}
\item[{$\bullet$}]  
The adiabatic perturbations come from $V_{\rm inf}$ and not from $V_{\rm int}$, that is, $V_{\rm inf} \gg V_{\rm int}$.

\item[{$\bullet$}] 
The inflaton dynamics is dominated by $\phi$, i.e., 
$\left|\frac{\partial V_{\rm inf}}{\partial \phi} \right|\gg \left|\frac{\partial V_{\rm int}}{\partial \phi}\right|$.

\item[{$\bullet$}] 
The quantum fluctuations of axions are not over-damped during the inflation, $m_{\theta}^2,m_{\phi}^2 \ll H^2$.

\item[{$\bullet$}] 
The standard slow-roll conditions, $\epsilon\ll 1$ and $|\eta|\ll 1$.
\end{description}
Some of the above conditions may be redundant depending on the parameter range of interest. In the light of these conditions, the cross-correlation parameter is bounded above by
\begin{align}
\beta_C \simeq 
4.2 \left( \frac{1}{g_1g_2} \right)
\left(
\frac{\Lambda_2^4}{A_S}
\right) \cos(\psi_0+\theta_0)
\frac{\phi_0^2}{96\pi^2} 
\ll 
2.1\left( \frac{1}{g_1g_2} \right),
\label{eq:betac2nat}
\end{align} 
where $|\Lambda_2^4 \cos(\psi_0+\theta_0)| \ll |V_{\rm inf}|$ is applied. 
For $g_1, g_2 <{\cal O}(1)$, the constraint of Eq.~(\ref{eq:betac2nat}) for $\beta_{\calC}$ 
is automatically satisfied if Eq.~(\ref{eq:betac1nat}) is satisfied. 
The illustrative values of isocurvature parameters are  
listed in Tab.~\ref{tab:1} by setting the typical values for the parameters in the scalar potential~in Eq. (\ref{eq:infponat3}). 
Note that, although the axion decay constants are typically of order the grand unification scale~($10^{16}$ GeV)~\cite{Choi:1985je} and hence one may expect $g_1\sim g_2$, the hierarchical values $g_1\ll g_2$ ($g_1\gg g_2$) can well be realized for the axion $\chi$ ($\phi$) by the non-perturbative effects through the (gauge) threshold correction~(Eq. (\ref{eq:threshold})). 

\begin{figure}[htb!]
\begin{center}    
  \psfrag{nRR}[B][B][1][0]{$n_s$}
  \psfrag{bbb}[B][B][1][0]{$\beta_{iso}$}
  \includegraphics[scale=0.4]{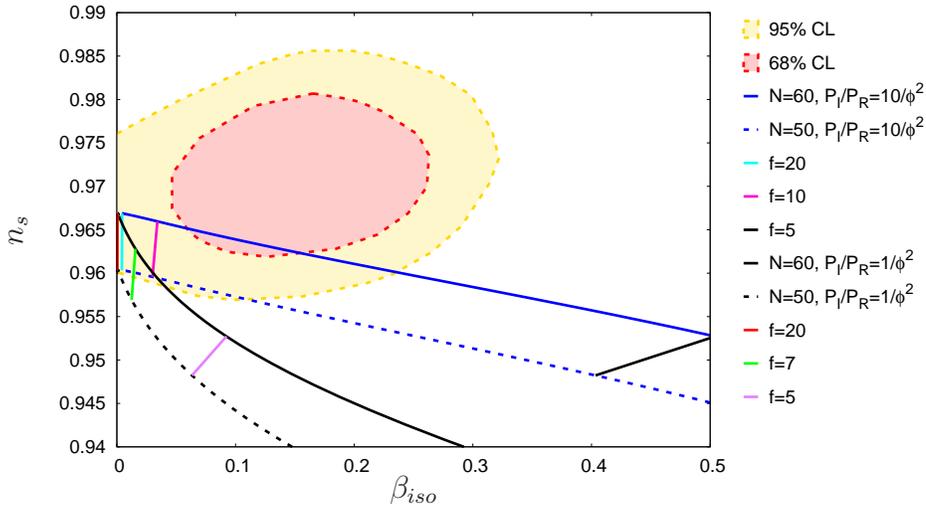} 
  \end{center}  
\caption{$\beta_{\rm iso}\equiv \calP_{\calI}/(\calP_{\calR}+\calP_{\calI})$ and $n_{s}$ for the natural inflation with sinusoidal correction ($68\%$ and $95\%$ CL contours are from Planck \cite{Ade:2015lrj}).   
$(\beta_{\rm iso}, n_s)$ are shown for $\calP_{\calI}/\calP_{\calR}=c/\phi_0^2$ with $c=1,10$ for varying $f$ and $N$. 
}
\label{crossisonat}
\end{figure}

Next, we estimate the fraction of isocurvature perturbations
\begin{align}
\beta_{\rm iso}=\frac{{\cal P}_{\cal I}}{{\cal P}_{\cal R}+{\cal P}_{\cal I}}
=\frac{ \frac{{\cal P}_{\cal I}}{{\cal P}_{\cal R}}}{ 1+\frac{{\cal P}_{\cal I}}{{\cal P}_{\cal R}}},
\end{align} 
where the power spectrum of the adiabatic perturbation is fixed as in Eq. (\ref{askn}), 
whereas the power spectrum of the 
isocurvature perturbation is given by 
\begin{align}
{\cal P}_{\cal I} \approx \left(\frac{\Omega_a}{\Omega_m}  \right)^2
\left( \frac{1}{2\pi} \right)^2  
\left( \frac{2}{g_1\theta_0} \right)^2 \frac{\Lambda_1^4}{6} \left( \frac{\phi_0^2}{f^2} \right) 
=
\left(\frac{\Omega_a}{\Omega_m}  \right)^2
\left( \frac{1}{g_1\theta_0} \right)^2 \frac{16A_S}{\phi_0^2}. 
\end{align} 
Then, the fraction of isocurvature perturbations
\begin{align}
\frac{{\cal P}_{\cal I}}{ {\cal P}_{\cal R}} \approx
16\left(\frac{\Omega_a}{\Omega_m}  \right)^2 
\left( \frac{1}{g_1\theta_0} \right)^2 \phi_0^{-2},
\label{eq:fraciso}
\end{align} 
can give a sizable contribution to the 
cosmological observables as illustrated in Fig.~\ref{crossisonat} where the prefactor of 
$\phi_0^{-2}$ in Eq.~(\ref{eq:fraciso}) is set to $1$ and $10$ for a varying $f$.
A larger prefactor is preferred for a larger isocurvature contributions.
The Planck bounds the uncorrelated axion isocurvature mode to $\beta_{\rm iso}\lesssim 0.038$, whereas the inclusion of isocurvature cross-correlation results in the constraint 
$0.034 \lesssim \beta_{\rm iso}\lesssim 0.28$ at the $95\%$ confidence level \cite{Ade:2015lrj}. 
Tab.~\ref{tab:1} summarizes the typical numerical values of parameters in the scalar potential~(Eq. (\ref{eq:infponat3})) 
which can realize a sizable fraction of isocurvature perturbations.

\begin{table}[htb]
\begin{center}
\begin{tabular}{|c|c|c|c|c|c|c|c|c|c|c|} \hline
$f$ & $N$ & $ g_1$ & $ g_2$ & $\Lambda_2^4/\Lambda_1^4$ & $\Omega_a/\Omega_m$ & $\cos(\psi_0+\theta_0)$ &  $\theta_0$ & $\beta_{\cal C}$ & $\beta_{\rm iso}$ & $n_{s}$\\ \hline
$10$ & $55$ & $10^{-4}$ & $10^{-2}$ & $2\times 10^{-10}$ & $2\times 10^{-4}$ & $1/2$ & $2$ & $2\times 10^{-7}$ & $0.07$ & $0.964$\\ \hline  
$10$ & $55$ & $10^{-2}$ & $10^{-2}$ & $1\times 10^{-5}$ & $0.02$ & $1/2$ & $2$ & $1\times 10^{-4}$ & $0.07$ & $0.964$\\ \hline  
$10$ & $55$ & $10^{-2}$ & $1$ & $1\times 10^{-5}$ & $0.02$ & $1/2$ & $2$ & $1\times 10^{-6}$ & $0.07$ & $0.964$\\ \hline  
$10$ & $55$ & $1$ & $10^{-2}$ & $1\times 10^{-5}$ & $0.02$ & $1/2$ & $2$ & $1\times 10^{-6}$ & $7\times 10^{-6}$ & $0.964$\\ \hline  
\end{tabular}
 \caption{The typical numerical values of parameters, the e-folding number ($N$), the spectral index ($n_{s}$), the fraction of isocurvature perturbation ($\beta_{\rm iso}$) 
and the cross-correlation parameter ($\beta_{\calC}$) for the natural inflation with sinusoidal correction.}
\label{tab:1}
\end{center}
\end{table}

Before concluding this section to move on to the discussion on the monodromy inflation, let us comment on the microscopic description about the correction term in 
the inflaton potential given by Eq. (\ref{eq:infponat2}). 
Such a potential can be derived from the following 
K\"ahler and superpotential,
\begin{align}
K=-2\ln (T_1+\bar{T}_1) -\ln (T_2+\bar{T}_2),
\nonumber\\
W=w_0 +Ae^{-b_1T_1} +Be^{-c_1T_1-c_2T_2},
\label{eq:KW}
\end{align}  
where $w_0$ is the flux-induced constant term induced by the Gukov-Vafa-Witten 
superpotential $W_{\rm flux}=\int G \wedge \Omega$, where $G$ is the linear combination 
of Ramond-Ramond and Neveu-Schwarz three-form fluxes 
and $\Omega$ is the period vector in the framework of type II superstring theory. 
The second and third terms in the superpotential~(\ref{eq:KW}) denote the non-perturbative 
effects, such as the gaugino condensation terms, D-brane instanton and world-sheet 
instanton effects. 

Let us define the moduli as 
\begin{align}
T_1=t_1+i a_1,\nonumber\\
T_2=t_2+i a_2,
\end{align}  
and assume that all the real parts of moduli $T_{1,2}$ are 
stabilized at their minima and sufficiently heavier than the remaining 
imaginary parts of $T_{1,2}$. 
Then, from the four-dimensional scalar potential based on $4$D ${\cal N}=1$ supergravity, 
\begin{align}
V=e^K(K^{I\bar{J}} D_I W D_{\bar J}\bar{W} -3|W|^2),
\end{align}  
where $K^{I\bar{J}}$ is the inverse of the K\"ahler metric 
$K_{I\bar{J}}=\partial^2 K/\partial \Phi^I\partial\bar{\Phi}^{\bar{J}}$, 
$D_{I}W=W_{I}+K_{I}W$, with 
$W_{I}=\partial W/\partial \Phi^I$ and 
$K_{I}=\partial K/\partial \Phi^I$, for $\Phi^I=T^1, T^2$. We can obtain the axion potential for $a_{1,2}$ 
by further assuming that some uplifting sector 
lifts up the scalar potential from the AdS vacuum to the dS one with a very small 
vacuum energy, 
\begin{align}
V\simeq \Lambda +\Lambda_1 {\rm cos}\left( \frac{\phi}{f}\right) 
+\Lambda_2 {\rm cos}\left( \frac{\phi}{g_1}+\frac{\chi}{g_2}\right),
\end{align}  
where $\Lambda \simeq -\Lambda_1-\Lambda_2 \simeq -\Lambda_1$ is a constant and $\Lambda_{1,2}$ 
depend on the vacuum expectation values of ${\rm Re}\,T_{1,2}$. 
The fields $\phi$ and $\chi$ are the canonically normalized axions.
The kinetic terms of $a_{1,2}$ are extracted from the second derivatives of the K\"ahler potential with respect to the moduli,
\begin{align}
K_{I\bar{J}}\partial \Phi^I\partial \bar{\Phi}^J &=K_{T_1\bar{T}_1} \partial T_1 \partial {\bar T}_1 
+K_{T_2\bar{T}_2} \partial T_2 \partial {\bar T}_2.
\end{align}  
As a result, the axion decay constants $f_{1}, g_{1,2}$ of the canonically normalized axions $\phi$ and $\chi$ 
are given by 
\begin{align}
f&=\frac{\sqrt{2K_{T_1\bar{T}_1}}}{b_1} =\frac{2}{b_1(T_1+\bar{T}_1)},
\nonumber\\
g_1&=\frac{\sqrt{2K_{T_1\bar{T}_1}}}{c_1} =\frac{2}{c_1(T_1+\bar{T}_1)},
\nonumber\\
g_2&=\frac{\sqrt{2K_{T_2\bar{T}_2}}}{c_2} =\frac{\sqrt{2}}{c_2(T_2+\bar{T}_2)}.
\end{align}


\section{Axion monodromy inflation with sinusoidal correction} 
\label{sec:3}
We now discuss the axion monodromy inflation which offers another popular axion inflation scenario in string theory. Axion monodromy inflation is a successful large-field inflation 
in which the inflaton can move around its configuration place 
on many cycles, and the field range of inflaton can be thus much larger than its fundamental 
period determined by the axion decay constant. 

The scalar potential for the axion monodromy inflation is represented by
\begin{align}  
{\cal L}=-\frac{1}{2}(\partial \phi)^2 -\mu_1^{4-p}\phi^p,
\label{eq:mono}
\end{align}
where $\phi$ is the axion originating from the higher-dimensional form 
fields,  $\mu_1$ represents the energy scale and $p$ is the fractional number which depends on the model 
in string theory ~\cite{Baumann}.

Let us consider the spacetime filling D$5$-brane in type IIB string theory~\cite{McAllister:2008hb}.  
The D$5$-brane wraps a certain internal two-cycle $\Sigma_2$ in the $6$D compact space 
in addition to the $4$D spacetime and its {\it Dirac-Born-Infeld} action is given by
\begin{align}  
S_{D5}=\frac{1}{(2\pi)^5 g_s (\alpha^\prime)^3} \int d^6 \sigma 
\sqrt{-{\rm det} (G_{ab} +B_{ab}) },
\end{align}
where $g_s$ is the string coupling, $\alpha^\prime$ is the regge-slope, 
$G_{ab}$, $a,b=0,1,2,3,4,5$ is the pullback of the metric of the target space, 
$B_{ab}$ is the Kalb-Ramond field whose extra-dimensional component corresponds 
to the axion $b=\int_{\Sigma_2} B_2$ where $B_2$ is the Kalb-Ramond two-form. 
We here do not consider the magnetic flux background. 

After carrying out the dimensional reduction, the axion potential can be 
extracted as 
\begin{align}  
V_{\rm eff}\simeq \frac{{\cal T}}{(2\pi)^5 g_s (\alpha^\prime)^2} \sqrt{l^4 +b^2},
\end{align}
where ${\cal T}$ and $l$ are some warp factors and the volume of two-cycle $\Sigma_2$ in string units ($\alpha^\prime=1$). 
For a large field value of the inflaton $b\gg l^2$, the potential reduces to a linear type, 
\begin{align}  
V_{\rm eff}\simeq \frac{{\cal T}}{(2\pi)^5 g_s (\alpha^\prime)^2} b. 
\end{align}
Then, the relevant Lagrangian of the inflaton is given by  
\begin{align}  
{\cal L}=-\frac{1}{2}(\partial \phi)^2 -\mu_1^{3}\phi,
\end{align}
where $\mu_1^{3} =\frac{{\cal T}}{f (2\pi)^5 g_s (\alpha^\prime)^2}$ with $f$ being the decay constant of the axion $\phi=b$. 
Furthermore, for the D$4$-brane in a nilmanifold (twisted torus) on type IIA string theory, 
the axion potential has the form of Eq. (\ref{eq:mono}) with $p=2/3$~\cite{Silverstein:2008sg}. 
When we consider the seven-branes~\cite{Palti:2014kza} or a four-form field strength~\cite{Kaloper:2008fb}, 
the axion monodromy inflation is that with $p=2$. 
The other types of axion monodromy inflation with $p=4/3,3$ can also be constructed by a coupling between 
NS-NS two-form and the Ramond-Ramond field strength~\cite{McAllister:2014mpa}.

As mentioned for the natural inflation, the axion, in general, can receive the non-perturbative 
effects associated with the gaugino condensation, D-brane instanton and world-sheet instanton, and the scalar potential receives the moduli-dependent correction including the mixing with 
another light axion $\chi$,
\begin{align}  
V= \mu_1^{4-p}\phi^p +\mu_2^4 {\rm cos}\left( \frac{\phi}{g_1}+\frac{\chi}{g_2}\right),
\label{eq:mono2}
\end{align}
where  $g_2$ denotes the decay constant of $\chi$. We here assume that the moduli except for the relevant axions under our discussion are fixed at their minimum 
and decoupled from our setup. 

For the notational brevity, we in the following define the parameters 
\begin{align}
\psi =\frac{\phi}{g_1},\,\,\,
\theta =\frac{\chi}{g_2},
\end{align} 
so that the total potential can be written as
\begin{align}
V= \mu_1^{4-p}\phi^p
+
\mu_2^4 
\cos\left(\psi+\theta\right).
\end{align} 

Analogously to the natural inflation discussed in the last section, the curvature and isocurvature perturbations in our monodromy inflation scenario read

\begin{align}
{\cal P}_{\cal R} & =A_Sk^{n_{s}-1},
\nonumber \\
{\cal P}_{\cal I} & \approx \left(\frac{\Omega_a}{\Omega_m}  \right)^2
\left( \frac{H}{2\pi} \right)^2 
\left( \frac{2}{g_1\theta_0} \right)^2 
\simeq \left(\frac{\Omega_a}{\Omega_m}  \right)^2
\left( \frac{2}{g_1\theta_0} \right)^2 
A_S 
\left(
\frac{p}{\phi_0}
\right)^2,
\label{eq:PRPImono}
\end{align}
with
\begin{align}
H^2&=\frac{\mu_1^{4-p}\phi_0^p}{3} =4\pi^2 A_S 
\left(
\frac{p}{\phi_0}
\right)^2,
\end{align} 
replacing $\mu_1$ by $A_S$ through the CMB normalization, 
$\mu_1^{4-p}=12\pi^2p^2 A_S\phi_0^{-p-2}$, 
where $p=2$ corresponds to the natural inflation case. 
The cross-correlation power spectrum then becomes
\begin{align}
{\cal P}_{\cal C} &\sim 
2.1 \left( \frac{1}{g_1 g_2} \right)
\left(
\frac{\mu_2}{\mu_1}
\right)^4
\left(
\frac{\mu_1}{\phi_0}
\right)^p
\cos(\psi_0+\theta_0)
\sqrt{{\cal P}_{\cal R}{\cal P}_{\cal I}} 
\nonumber\\
&=
2.1 \left( \frac{1}{g_1 g_2} \right)
\left(
\frac{\mu_2^4}{12\pi^2 p^2 A_S}
\right)
\phi_0^2
\cos(\psi_0+\theta_0)
\sqrt{{\cal P}_{\cal R}{\cal P}_{\cal I}}.
\end{align}

\begin{figure}
\begin{center}    
    \psfrag{nRR}[B][B][1][0]{$n_s$}
  \psfrag{PRI/Sqrt[PRR*PII]}[B][B][1][0]{$\calP_{\calC}/\sqrt{\calP_\calR\calP_\calI}$}
    \includegraphics[scale=0.4]{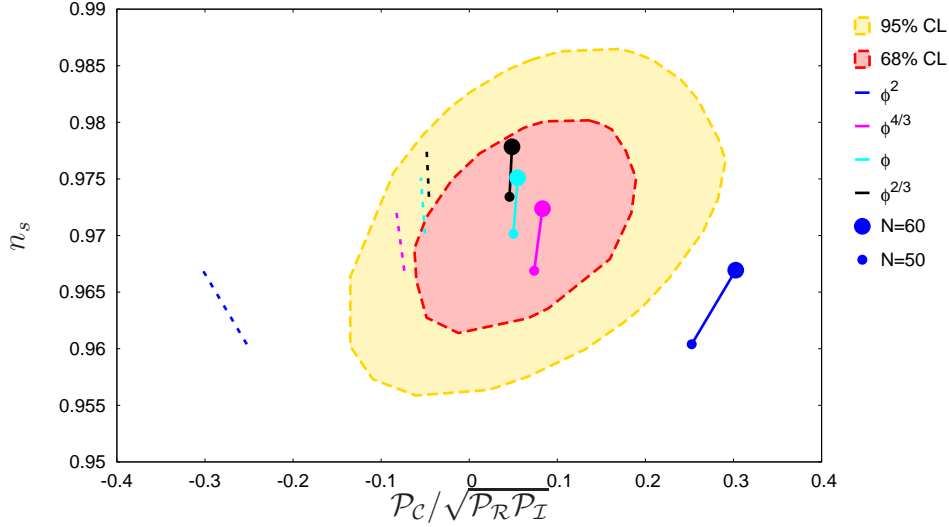} 
\end{center}
\caption{ $\calP_{\calC}/\sqrt{\calP_{\calR}\calP_{\calI}}$ and $n_{s}$ for the axion monodromy inflation ($V_{\rm inf}=\mu_1^{4-p}\phi^p$) with sinusoidal correction (68 \% and 95 \% CL contours are from Planck \cite{Ade:2015lrj}).
$\calP_{\calC}/\sqrt{\calP_{\calR}\calP_{\calC}}= c \times \phi_0^2/p^2$ for $c=0.005$ are shown for varying the e-folding number $N$. 
The anti-correlation cases (for $c=-0.005$) are also shown with the dashed lines.} 

\label{crosscmono}
\end{figure}

We plot the cross-correlation parameter
\begin{align}
\beta_C =\frac{{\cal P}_{\cal C}}{\sqrt{{\cal P}_{\cal R}{\cal P}_{\cal I}}} &=
2.1 \left( \frac{1}{g_1 g_2} \right)
\left(
\frac{\mu_2^4}{12\pi^2 A_S}
\right)
\cos(\psi_0+\theta_0) 
\left( \frac{\phi_0}{p}\right)^2,
\end{align} 
by varying the index $p$ and the e-folding number $N$ in Fig.~\ref{crosscmono}, where the prefactor of $\phi_0^2/p^2$ is set to $\pm 0.005$ for concreteness. $\beta_{\cal C}$ increases for a larger $p$ because the initial field value of axion-inflaton increases for a larger $p$. 
By considering the consistency conditions as spelled out below Eq. (\ref{eq:betac1nat}) in the last section for the validity of our calculations, 
the cross-correlation is bounded above by
\begin{align}
\beta_C 
&\simeq
2.1 \left( \frac{1}{g_1 g_2} \right)
\left(
\frac{\mu_2^4}{12\pi^2 p^2 A_S}
\right)
\phi_0^2
\cos(\psi_0+\theta_0)
\nonumber\\
&\ll 
2.1  \left( \frac{1}{g_1g_2} \right).
\label{eq:betac2}
\end{align} 
Tab.~\ref{tab:2} lists some illustrative values for the isocurvature perturbation parameters by setting the typical values for parameter sets in the scalar potential~(\ref{eq:mono2}).

\begin{figure}
\begin{center}    
    \psfrag{nRR}[B][B][1][0]{$n_s$}
  \psfrag{bbb}[B][B][1][0]{$\beta_{iso}$}
  \includegraphics[scale=0.4]{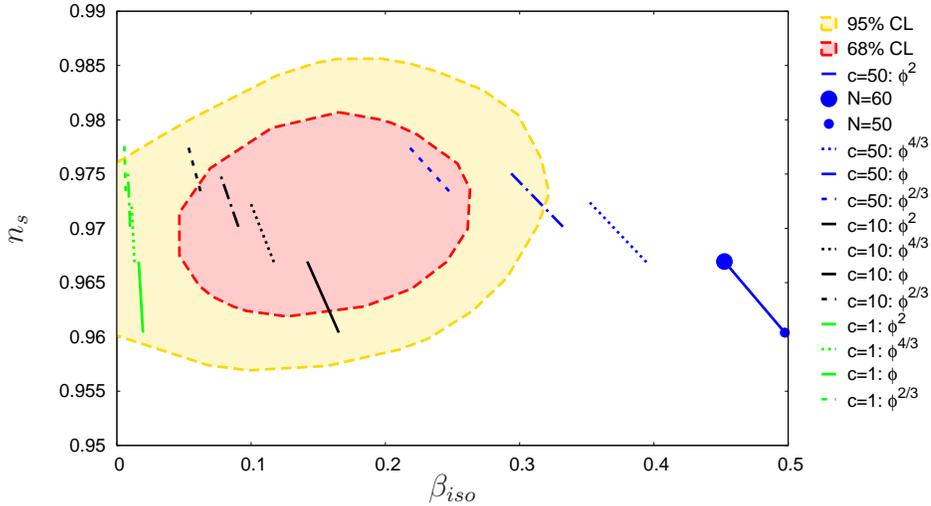} 
\end{center}  
\caption{$\beta_{\rm iso}\equiv \calP_{\calI}/(\calP_{\calR}+\calP_{\calI})$ and $n_{s}$ for the axion monodromy inflation ($V_{\rm inf}=\mu_1^{4-p}\phi^p$) 
with sinusoidal correction (68 \% and 95 \% CL contours from Planck \cite{Ade:2015lrj}). 
($\beta_{\rm iso}, n_s$) are shown for $\calP_{\calI}/\calP_{\calR}=c\times p^2/\phi_0^2$ with $c=1,5,50$.}

\label{crossisomono}
\end{figure}

We next estimate the fraction of isocurvature perturbations
\begin{align}
\beta_{\rm iso}=\frac{{\cal P}_{\cal I}}{{\cal P}_{\cal R}+{\cal P}_{\cal I}} 
=\frac{ \frac{{\cal P}_{\cal I}}{{\cal P}_{\cal R}}}{ 1+\frac{{\cal P}_{\cal I}}{{\cal P}_{\cal R}}},
\end{align} 
where the power spectrum of adiabatic perturbations is fixed to be $A_S=2.2 \times 10^{-9}$ at the pivot scale $k_{\ast}=0.05\,{\rm Mpc}^{-1}$, 
whereas the fraction of ${\cal P}_{\cal R}$ and ${\cal P}_{\cal I}$ is given by
\begin{align}
\frac{{\cal P}_{\cal I}}{{\cal P}_{\cal R}} 
\approx
\left(\frac{\Omega_a}{\Omega_m}  \right)^2
\left( \frac{2}{g_1\theta_0} \right)^2 
\left( \frac{p}{\phi_0} \right)^2.
\end{align} 
Fig.~\ref{crossisomono} plots $\beta_{\rm iso}$ with $p=2/3,1,4/3,2$ as 
a function $\phi_0$ by setting the parameters as 
\begin{align}
\frac{{\cal P}_{\cal I}}{{\cal P}_{\cal R}} =c \times \left( \frac{p}{\phi_0} \right)^2,
\end{align}  
with $c=1,10,50$, and we can find, as expected, the isocurvature contribution increases for a larger $p$. The Planck data hence favors the sizable generally correlated isocurvature perturbations for the axion monodromy inflation with sinusoidal correction. 
Tab.~\ref{tab:2} exemplifies the parameters which can realize the sizable fraction of isocurvature perturbations. Figs.~\ref{nsvsr},~\ref{crosscmono} and~\ref{crossisomono} hence demonstrates that, for the axion monodromy 
inflation with $p=1,2/3$ including the sinusoidal correction, there is a preference for the existence of cross-correlated isocurvature modes in the currently available CMB data. 

\begin{table}[htb]
\begin{center}
\begin{tabular}{|c|c|c|c|c|c|c|c|c|c|c|} \hline
$p$ & $N$ & $ g_1$ & $ g_2$ & $\mu_2^{4-p}/H^2$ & $\Omega_a/\Omega_m$ & $\cos(\psi_0+\theta_0)$ & $\theta_0$ & $\beta_{\cal C}$ & $\beta_{\rm iso}$ & $n_{s}$\\ \hline
$2$ & $55$ & $10^{-2}$ & $10^{-2}$ & $6\times 10^{-7}$ & $0.03$ & $1/2$ & $2$ & $0.002$ & $0.14$ & $0.964$\\ \hline  
$4/3$ & $55$ & $10^{-2}$ & $10^{-2}$ & $3\times 10^{-7}$ & $0.03$ & $1/2$ &$2$ & $0.001$ & $0.1$ & $0.97$\\ \hline  
$1$ & $55$ & $10^{-2}$ & $10^{-2}$ & $4\times 10^{-7}$ & $0.03$ & $1/2$ & $2$ &$0.001$ & $0.08$ & $0.973$\\ \hline  
$2/3$ & $55$ & $10^{-2}$ & $10^{-2}$ & $4\times 10^{-7}$ & $0.03$ & $1/2$ & $2$ & $0.001$ & $0.05$ & $0.976$\\ \hline  
\end{tabular}
\caption{The typical numerical values for the axion monodromy inflation with sinusoidal correction.}
\label{tab:2}
\end{center}
\end{table}

\clearpage
\section{Conclusion}
\label{sec:con}
The natural inflation and axion monodromy inflation with moduli-dependent sinusoidal corrections can generically appear in the 
scalar potential through the non-perturbative effects. We demonstrated that probing the precise nature of isocurvature fluctuations can help us understand the nature of fundamental physics using these popular inflation models as concrete examples. 
In this paper, we focused on the scenarios where the heavy axion induces the adiabatic perturbations while another light axion sources the isocurvature 
perturbations with their cross-correlations taken into account. 

Sec.~\ref{sec:2} demonstrated that the cross-correlated isocurvature mode gives an even tighter constraints on the decay constant of axion-inflaton and the e-folding number for natural inflation.
 While Sec.~\ref{sec:2} showed that the cross-correlated isocurvature perturbations are not favored by Planck for the natural inflation, Sec.~\ref{sec:3} showed that there is a preference for the existence of cross-correlated isocurvature mode for the axion monodromy inflation.


We also mention that, when, in contrast to our setup, the sinusoidal corrections are not suppressed enough, the scalar power spectrum could posses the modulating behaviour for the natural inflation~\cite{Abe:2014xja,Czerny:2014wua} and the axion monodromy inflation~\cite{Flauger:2009ab,Kobayashi:2010pz,Kobayashi:2014ooa,Higaki:2014sja}. Such an additional feature in the cosmological observables would also be of great interest to discriminate among the possible inflation models.

We illustrated our findings through the simple models where a single axion is added besides the axion-inflaton, but, in general, there could appear multiple light axions $\chi_i$ in addition to 
the axion-inflaton $\phi$. They can then have the mixing terms 
\begin{eqnarray}
\sum_j \Lambda^4_j \left( 1- \cos \left( \frac{\phi}{g} +\sum_i \frac{\chi_i}{g_i^j}  \right) \right),
\end{eqnarray}
and the analysis analogous to what has been done in this paper can be preformed for such multiple axion cases too.

Even though Planck, in particular the addition of the polarization data, can significantly tighten the constraints on the isocurvature modes, 
we should be careful about the potential systematics in the current Planck data. 
For instance, it was pointed out that the apparent low power in $TE$ spectrum could result in the preference for 
the positive cross-correlation, and this could cause an over-constraint on the isocurvature component if 
such a power spectrum feature is due to the unidentified systematics~\cite{Ade:2015xua}. 
While the current polarization data at hand are not yet robust to the systematis on the large scales, 
the forthcoming polarization data with a better handle on the systematics would certainly be able to probe the nature of isocurvature perturbations more precisely 
and consequently explore the properties of ubiquitous light degrees of freedom in the early Universe.

\subsection*{Acknowledgements}
We thank Jinn-ouk Gong for the useful discussions and the hospitality of CTPU
where the collaboration initiated. 
K.~K. was supported by Institute for Basic Science (IBS-R018-D1) and thanks the Galileo Galilei Institute for Theoretical Physics for hospitality. 
T.~K. was supported in part by
the Grant-in-Aid for Scientific Research No.~25400252 and No.~26247042 from the 
Ministry of Education,
Culture, Sports, Science and Technology (MEXT) in
Japan. 
H.~O. was supported in part by a Grant-in-Aid for JSPS Fellows 
No. 26-7296.

\end{document}